\documentclass{iaus}
\pdfoutput=1
\usepackage[utf8]{inputenc}
\usepackage[english]{babel}
\usepackage{graphicx}
\usepackage{astron}

\newcommand{\pc}{\,\textrm{pc}}
\newcommand{\kpc}{\,\textrm{kpc}}
\newcommand{\yr}{\,\textrm{yr}}
\newcommand{\Myr}{\,\textrm{Myr}}
\newcommand{\Gyr}{\,\textrm{Gyr}}

\newcommand{\erg}{\,\textrm{erg}}

\newcommand{\muG}{\,\mu{\textrm{G}}}

\newcommand{\apj}{ApJ}
\newcommand{\apjs}{ApJS}
\newcommand{\apjl}{ApJL}

\newcommand{\aap}{A\&A}

\title{COSMIC--RAY DRIVEN DYNAMO IN GALAXIES}

\author[M.Hanasz et al.] {M. Hanasz$^1$, D. Wóltanski$^1$, K. Kowalik$^1$ \and H. Kotarba$^2$}

\affiliation{$^1$Centre for Astronomy, Nicolaus Copernicus University, ul. Gagarina 11, PL-87-100 Torun, Poland\break
   $^1$University Observatory Munich, Scheinerstr. 1, D-81679 Munich, Germany}

\pubyear{2010}
\volume{274} 
\pagerange{???--???}
\jname{Advances in Plasma Astrophysics}
\editors{A. Bonanno, E. de Gouveia Dal Pino \& A. Kosovichev, eds.}
\begin{document}
\maketitle
\begin{abstract}
We present recent developments of  global galactic-scale numerical models of the Cosmic Ray (CR) driven dynamo, which was originally proposed by Parker (1992). We conduct a series of direct CR+MHD numerical simulations of the dynamics of the interstellar medium (ISM), composed of gas, magnetic fields and CR components. We take into account CRs accelerated in randomly distributed supernova (SN) remnants, and assume that SNe deposit small-scale, randomly oriented, dipolar magnetic fields into the ISM. The amplification timescale of the large-scale magnetic field resulting from the CR-driven dynamo is comparable to the galactic rotation period. The process efficiently converts small-scale magnetic fields of SN-remnants into galactic-scale magnetic fields. The resulting magnetic field structure resembles the X-shaped magnetic fields observed in edge-on galaxies.
\end{abstract}
\firstsection %
%
%
%
\section{Introduction}
The dynamical role of CRs was first recognized by Parker~\cite*{1966ApJ...145..811P}, who noticed that a vertically stratified ISM which consists of
thermal gas, magnetic fields and CRs is unstable due to buoyancy of the weightless components, i.e. the magnetic fields and the CRs. According to diffusive shock acceleration models CRs are continuously supplied to the ISM by SN remnants. Therefore, the buoyancy effects caused by CRs are expected in all star
forming galaxies. Theories of diffusive shock acceleration predict that about 10 \% of the $\sim 10^{51} \erg$ of the SN II explosion energy is
converted to CR energy. Observational data indicate that gas, magnetic fields and CRs appear in approximate energetic equipartition, which means
that all three components are dynamically coupled. In order to incorporate the CR propagation in MHD considerations we use the diffusion--advection equation
(e.g. Schlickeiser~\&~Lerche~\cite*{1985A&A...151..151S}) and take into account the CR pressure gradient in the gas equation of motion
(see e.g. Berezinski~et~al.~\cite*{1990acr..book.....B}).
\par The CR-driven dynamo was originally proposed by Parker \cite*{1992ApJ...401..137P}. Our model of the CR-driven dynamo involves the following
elements~\cite{2004ApJ...605L..33H,2006AN....327..469H,2009A&A...498..335H,2009ApJ...706L.155H}:
(1) The CR nuclear component  described by the diffusion--advection transport equation, supplemented to the standard set of
resistive MHD equations \cite{2003A&A...412..331H}.
(2) CRs supplied in SN remnants. The CR input of individual SNe is assumed to be $10\%$ of the typical SN kinetic energy output
($=10^{51}{\erg}$), while the thermal energy output from supernovae is neglected.
(3) Anisotropic CR diffusion along magnetic field lines~\cite{1999ApJ...520..204G},
(4) Finite resistivity of the ISM in order to permit topological evolution of the galactic magnetic fields via anomalous resistivity processes~\cite{2002A&A...386..347H},
and/or via turbulent reconnection~\cite{2009ApJ...700...63K} on small spatial scales, which are unresolved in our simulations.
(5) An initial gas distribution in the disk which follows the model of the ISM in the Milky Way by Ferri\`ere~\cite*{1998ApJ...497..759F}.
(6) Differential rotation of the interstellar gas, which currently follows an assumed form of a galactic gravitational potential.
\par We briefly mention that various properties of the shearing-box models of the CR-driven dynamo were discussed in a series of papers.
Computations of the dynamo coefficients in Parker unstable disks with CRs and shear are described by Kowal et al. \cite*{2006A&A...445..915K}
and by Otmianowska-Mazur et al. \cite*{2007ApJ...668..110O}.  Synthetic radio-maps of a global galactic disk based on local CR-driven dynamo models
exhibiting  X-type structures were presented by Otmianowska-Mazur et al. \cite{2009ApJ...693....1O}.
More recently, Siejkowski et al. \cite*{2010A&A...510A..97S} (see also Siejkowski et al. 2010, this volume) demonstrated that the CR-driven dynamo can also work
given the physical conditions of irregular galaxies, characterized by a relatively weak rotation and shearing rate.

\section{Global CR-driven dynamo simulations}
\begin{figure} 
 \centerline{ \includegraphics[width=0.46\columnwidth]{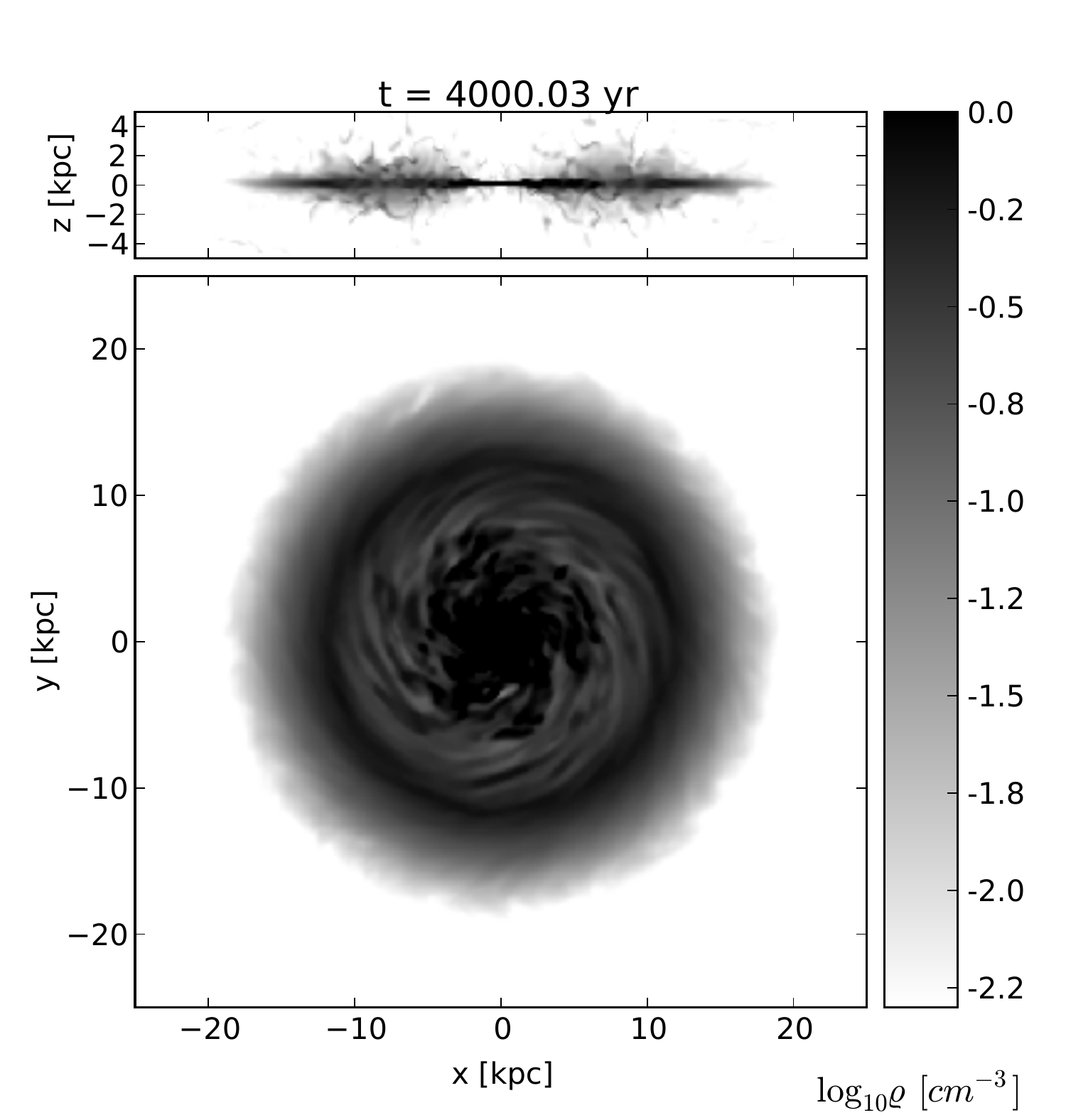}\quad \quad \includegraphics[width=0.47\columnwidth]{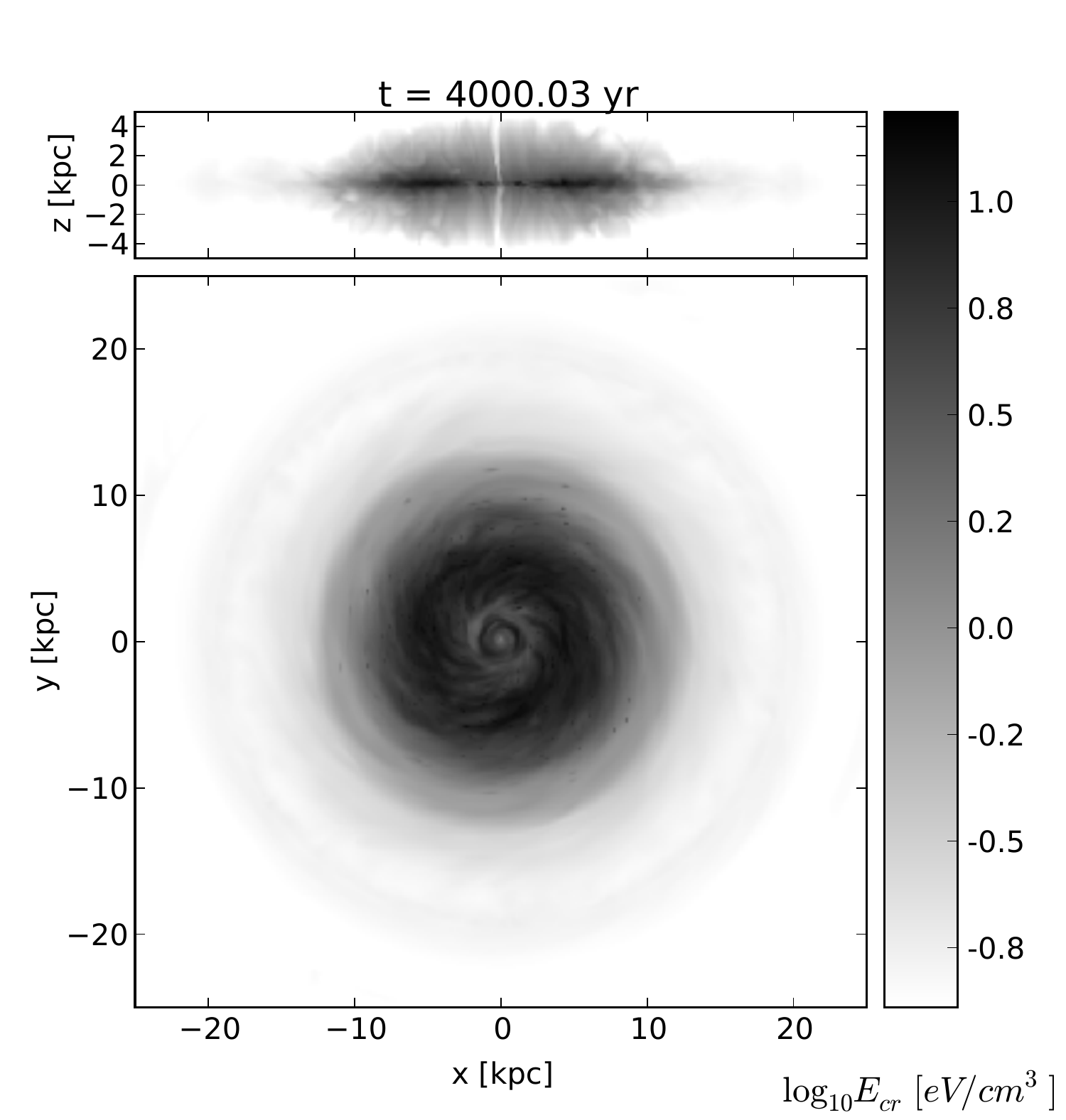}}
\centerline{ \includegraphics[width=0.50\columnwidth]{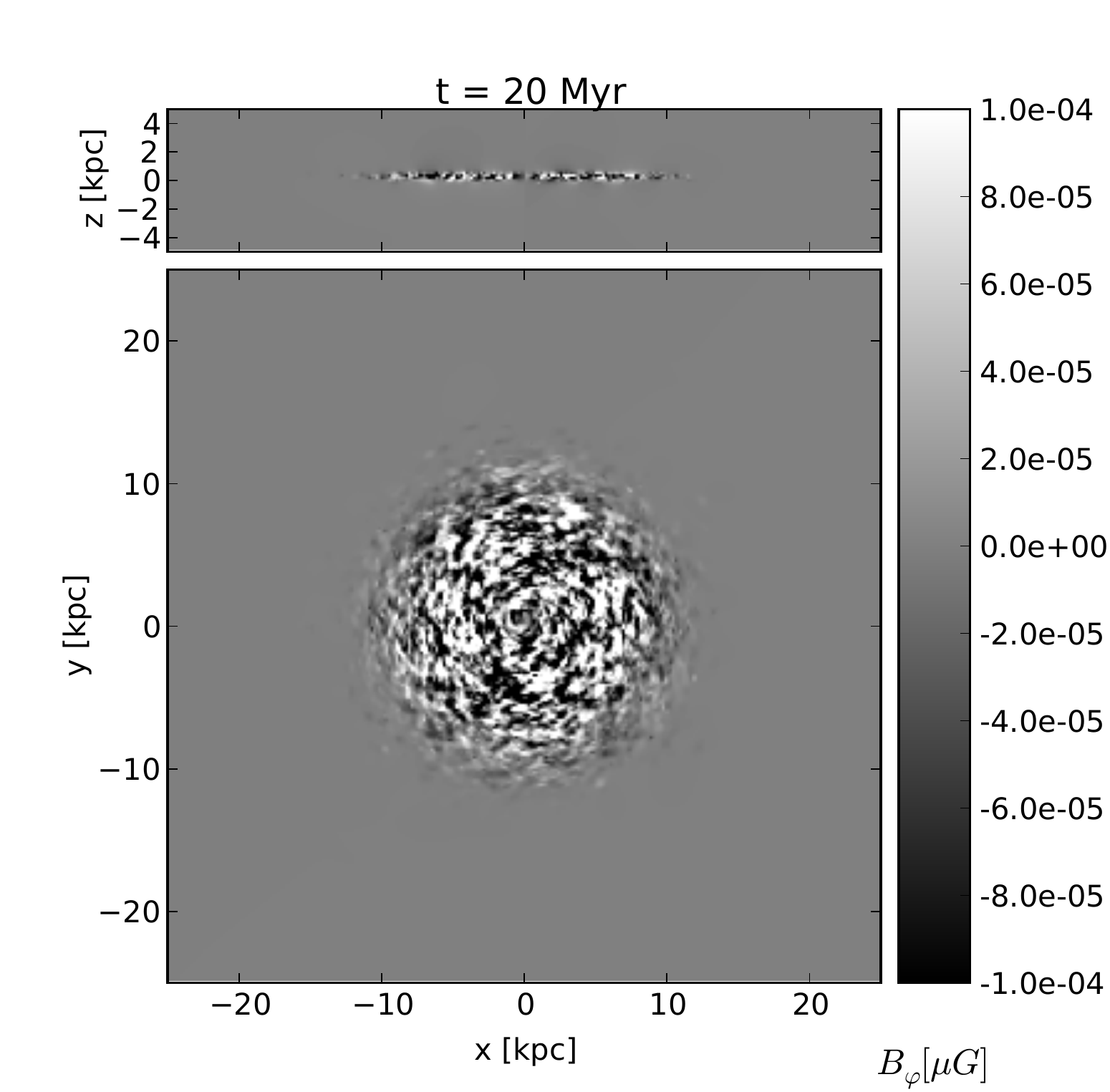} \includegraphics[width=0.50\columnwidth]{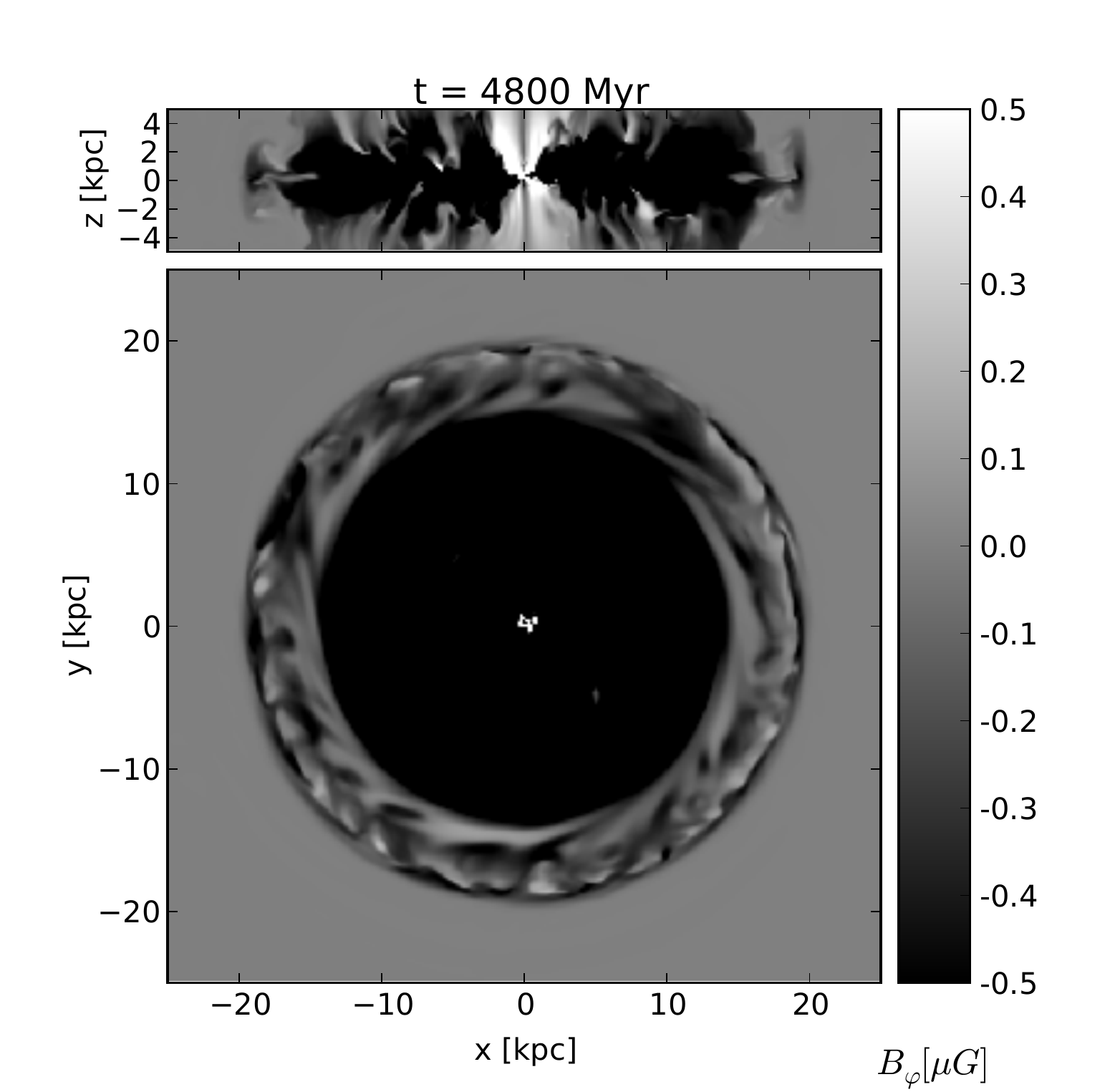}}
\caption{Top panels: logarithm of gas number density (left) and cosmic ray energy density (right)  at $t= 4\Gyr$. Bottom panels:
the distribution of toroidal magnetic field at $t=20 \Myr$ (left) and $t =4.8\Gyr$ (right). Unmagnetized regions of the volume are grey, while positive and negative toroidal magnetic fields are marked lighter  and darker, respectively. Note that the grey-scale scale in magnetic field maps is saturated to enhance weaker magnetic field structures in disk peripheries.}
\label{fig:d-cr-b-fixedg}
\end{figure}
\begin{figure}
\centerline{ \includegraphics[width=0.36\columnwidth]{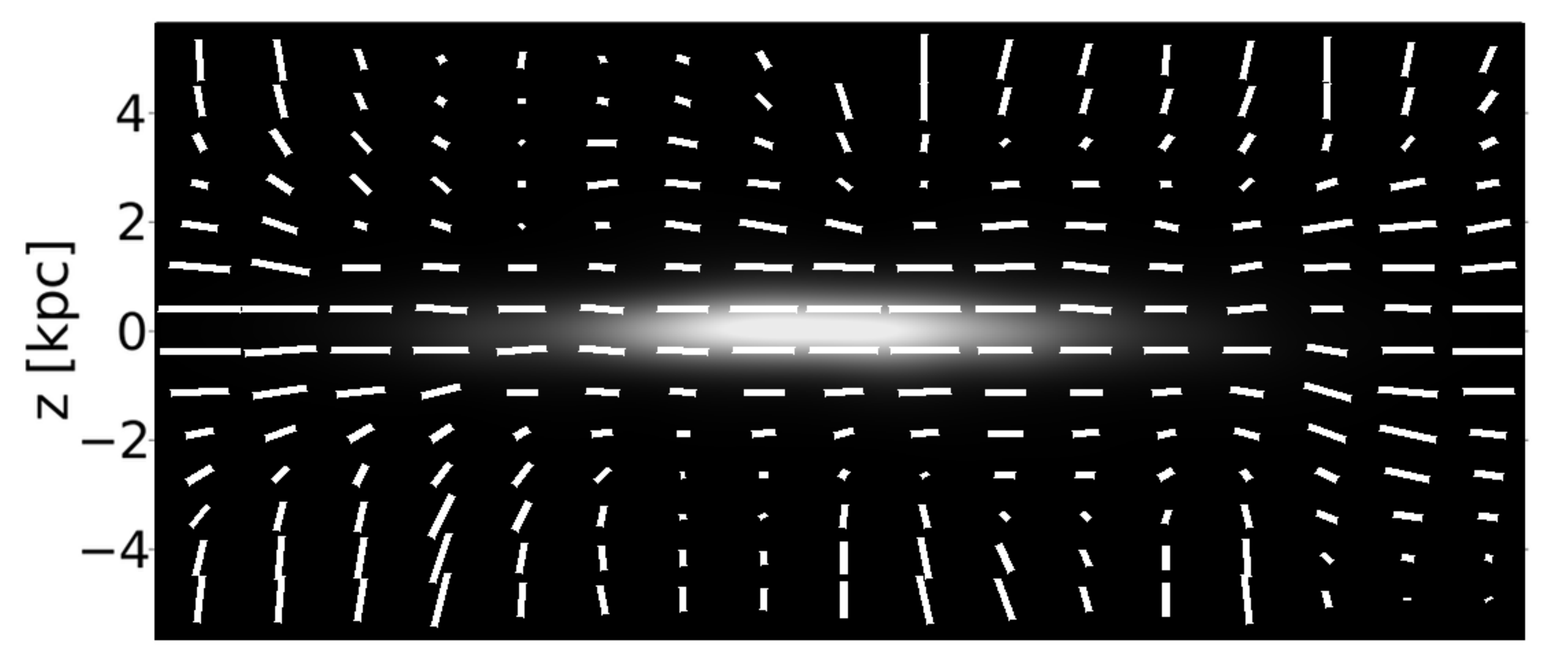}}
\centerline{ \includegraphics[width=0.36\columnwidth]{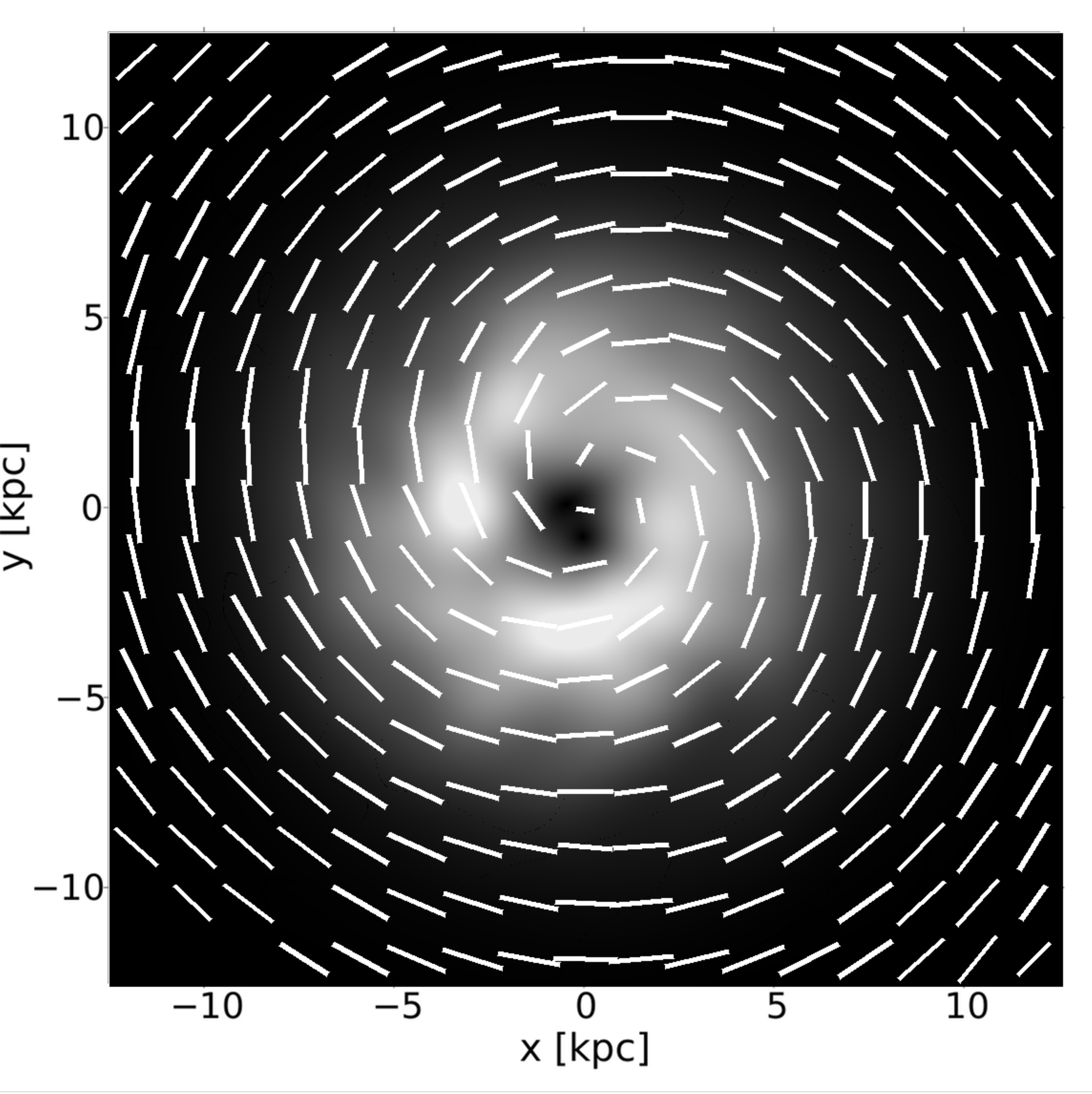}}
\caption{Synthetic radio maps of polarized intensity (PI) of synchrotron emission, together with polarization vectors are shown for the edge-on and face-on views of the galaxy at $t=4.8{\Myr}$.
Vectors direction resembles electric vectors rotated by $90^{\circ}$, and their lengths are proportional to the degree of polarization.}
\label{fig:radiomaps}
\end{figure}
Our first realization of a global CR+MHD galactic disk model relies on the following assumptions: First of all, we adopt analytical formulae for the
gravitational potential corresponding to a system consisting of a galactic halo, bulge and disk \cite{1991RMxAA..22..255A}. We assume no magnetic
field at $t=0$, and that weak ($10^{-4} \muG$), dipolar, small scale ($r \sim 50 \pc)$, randomly oriented magnetic fields are supplied locally in 10\% of the SN remnants for $t \leq 1\Gyr$.
We assume also that the SN rate is proportional to the star formation rate (SFR) which, on the other hand, is proportional to the initial gas column density.

\par We use the PIERNIK MHD code~\cite{2010EAS....42..275H,2010EAS....42..281H,2008arXiv0812.4839H,2009arXiv0901.0104H}, which is a grid-MHD code
based on the Relaxing TVD (RTVD) scheme~\cite{jin-xin-95,2003ApJS..149..447P}. PIERNIK is parallelized by means of block decomposition with the aid of
the MPI library. The original scheme was extended to deal with dynamically independent but interacting fluids, i.e. thermal gas and a diffusive CR gas,
which is described within the fluid approximation~\cite{2008arXiv0812.4839H}.
\par We find magnetic field amplification originating from the small-scale, randomly oriented dipolar magnetic fields, which is apparent through the
exponential growth by several orders of magnitude of both the magnetic flux and the magnetic energy (details see Hanasz et al. \cite*{2009ApJ...706L.155H}).
The growth phase of the magnetic field starts at the beginning of the simulation. 
The growth of the magnetic field strength saturates at about $t = 4\Gyr$, reaching values of $3-5{\muG}$ in the disk. During the
amplification phase, magnetic flux and total magnetic energy grow by about 6 and 10 orders of magnitude, respectively. The average e-folding time of
magnetic flux amplification is approximately equal to $270{\Myr}$, corresponding to the rotation at the galactocentric radius ($\approx10{\kpc}$). The magnetic
field is initially entirely random ($t=20{\Myr}$), since it originates from randomly oriented magnetic dipoles. Later on, the toroidal magnetic field
component forms a spiral structure revealing reversals in the plane of the disk. The magnetic field structure evolves gradually towards larger and lager scales.
The toroidal magnetic field component becomes almost uniform inside the disk at $t=2.5\Gyr$. The volume occupied by the well-ordered magnetic field
expands continuously until the end of the simulation.
\par In order to visualize the magnetic field structure in a manner resembling radio observations of external galaxies, we construct synthetic radio maps of the
synchrotron radio-emission, assuming that energy density of CR electrons equals 1\%  energy density of CR nucleons.
We apply standard procedures of line-of-sight integration of the stokes parameters {\it I},
{\it Q}, and {\it U} for the polarized synchrotron emissivity. We neglect the effects of Faraday rotation.
\par In Figure~\ref{fig:radiomaps}, we show the polarized intensity of synchrotron emission (grey-scale maps), together with polarization vectors. Electric
vectors, computed on the basis of integrated Stokes parameters, are rotated by $90^{\circ}$ to reproduce the magnetic field direction averaged along
the line-of-sight, assuming vanishing Faraday rotation effects. The polarization vectors, indicating the mean magnetic field direction,
reveal a regular spiral structure in the face-on view, and the so-called \textit{X-shaped structure} in the edge-on view. A particular similarity can
be noticed between our edge-on synthetic radio map and the radio maps of observed edge-on galaxies such as NGC 891 \cite{2009RMxAC..36...25K}.
In the present global model, the X-shaped configuration is an intrinsic property of the magnetic field structure, since it corresponds closely to the
flaring radial distribution of magnetic field in the disk and its neighborhood, as shown in Figure~\ref{fig:d-cr-b-fixedg}.
\par The face-on synchrotron radio map reveals a spiral structure of the magnetic field, however, due to the assumed axisymmetric gravitational
potential no features resembling spiral magnetic arms are present. To make the model more realistic, we incorporate non-axisymmetry in the
gravitational potential, applying two different approaches. The first approach relies on the addition of an analytical elliptical perturbation to the axisymmetric gravitational
potential. The results of this approach, which are presented by Kulpa-Dybel et al. 2010 (this volume), indicate that the CR-driven dynamo model reveals new properties,
such as the presence of a ring-like structure as well as a shift of the magnetic arms with respect to the crests of spiral density waves.
\par
\begin{figure}
\centerline{\hspace{-2mm}\includegraphics[width=0.43\columnwidth]{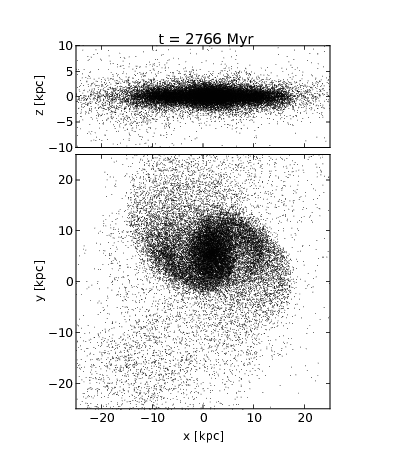}\qquad\includegraphics[width=0.43\columnwidth]{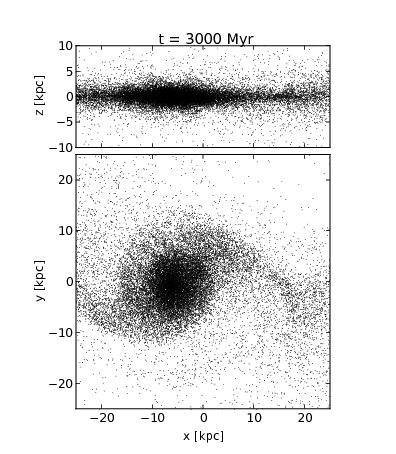}}
\caption{Two states of the N-body galactic disk prior to the galactic merger. The companion galaxy is apparent in the right panels.}
\label{fig:n-body}
\end{figure}

\begin{figure}
\centerline{\includegraphics[width=0.42\columnwidth]{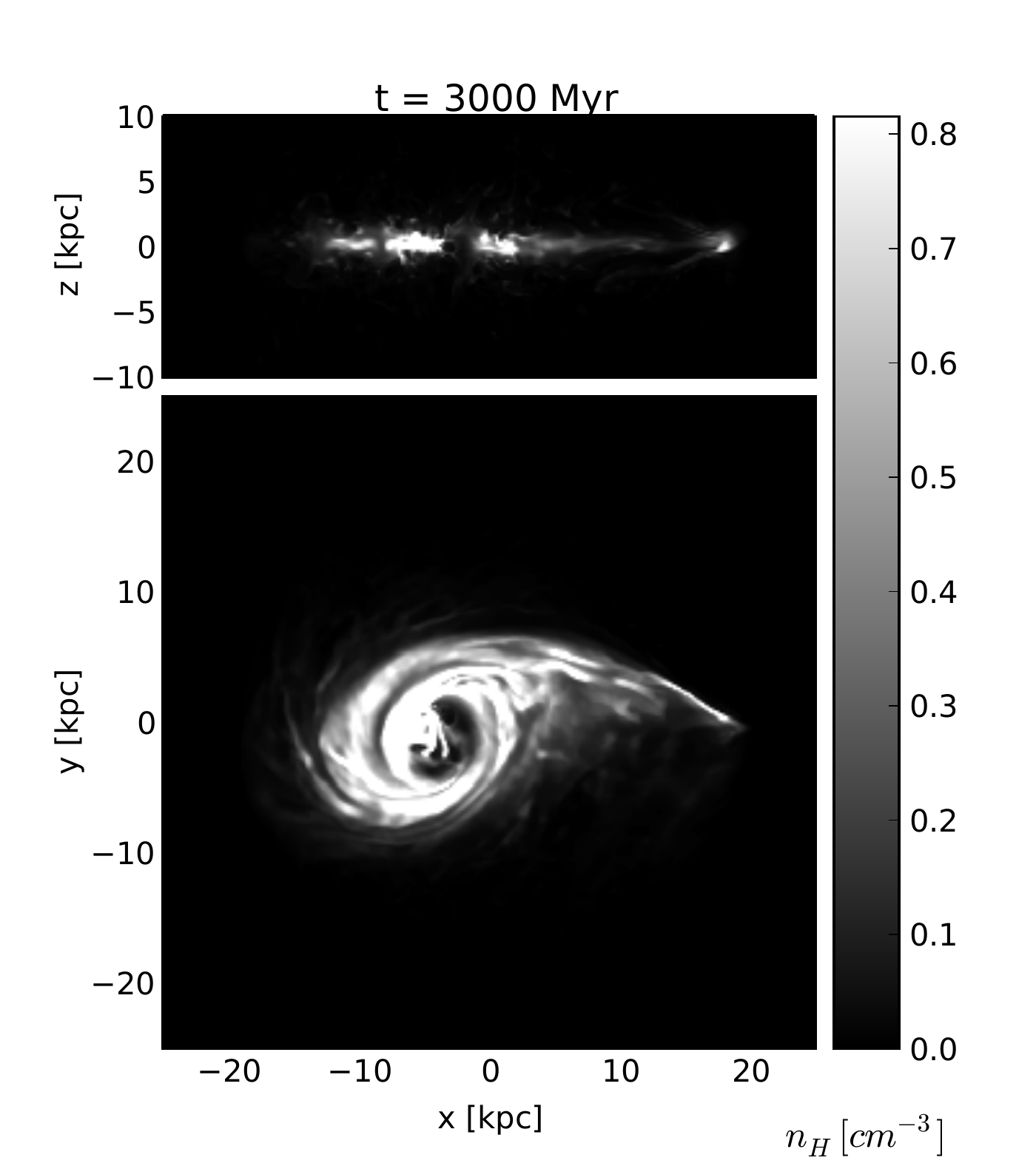}\quad\quad\quad\includegraphics[width=0.42\columnwidth]{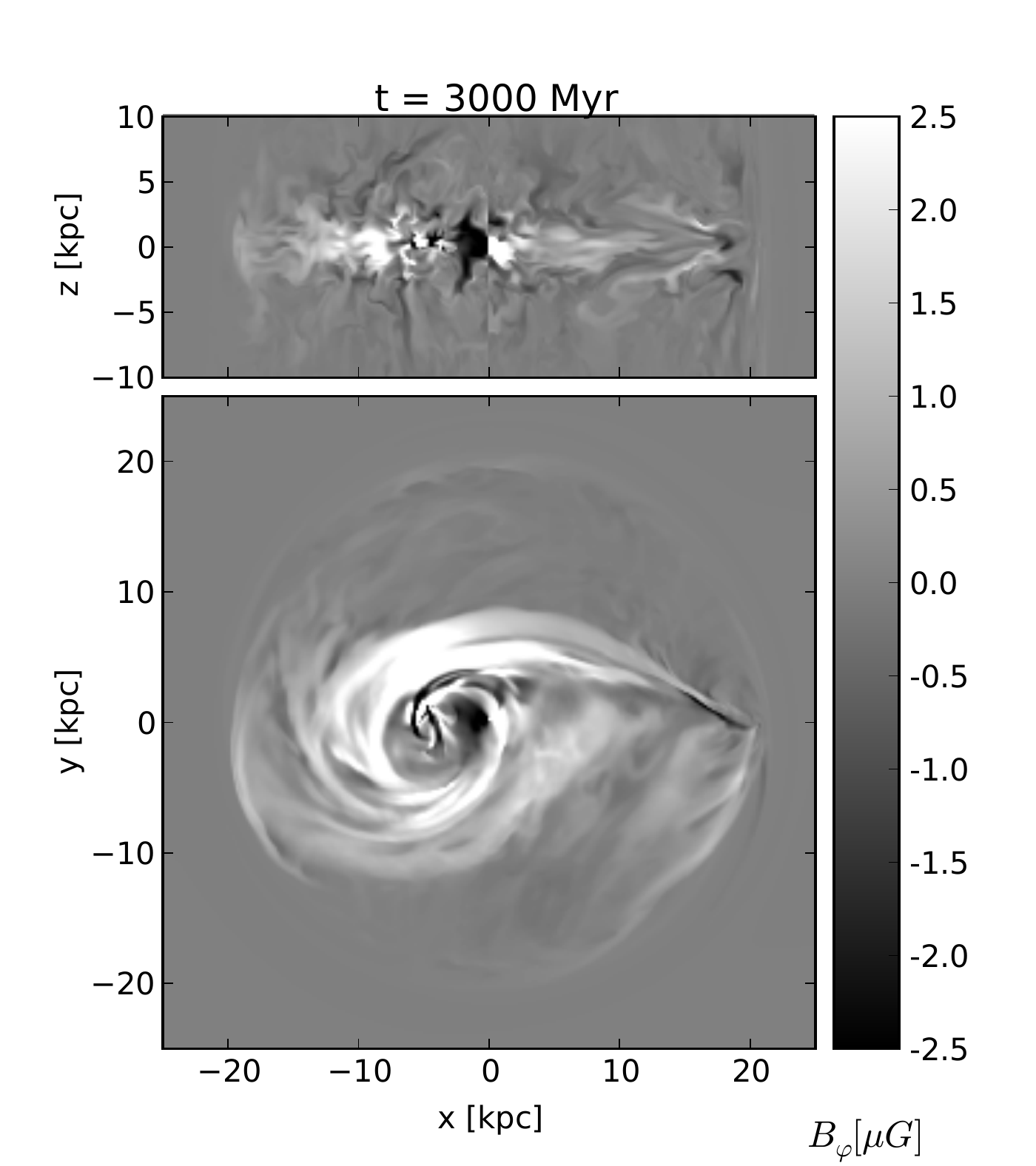}}
\caption{Gas density  (left panel) and magnetic field  (right panel) corresponding to the gravitational potential of the N-body system at the right plot of Fig.\ref{fig:n-body}}
\label{fig:dens-btor-n-body}
\end{figure}

 As a further step towards  more realistic galactic magnetic field models we perform N-body simulations of a disk-bulge-halo
system~\cite{2011ApJ-woltanski-etal}, and interpolate the resulting gravitational potential onto the computational grid. We use this potential to compute the gravitational
acceleration acting on the fluid components. In order to excite density waves in the galactic disk we add a small satellite galaxy, which ultimately merges
with the main galaxy at $t=3.2\Gyr$. Two snapshots of the N-body disk simulation are displayed in  Fig.~\ref{fig:n-body}.
The N-body part of the computation is performed with the VINE code \cite{2009ApJS..184..298W}, and the CR+MHD part with the PIERNIK code.
Fig.~\ref{fig:dens-btor-n-body} shows the gas density and the toroidal magnetic field component. Similar to the case of an
axisymmetric gravitational potential we observe efficient magnetic field amplification by the CR-driven dynamo in the presence of density waves in the
galactic disk. In the presence of spiral arms in the stellar and gas components the magnetic field also reveals a spiral structure. Moreover, we notice that both polarities  of the azimuthal magnetic field are present at an advanced stage of magnetic field evolution. It appears that
during the merger phase and afterwards the magnetic field structure becomes even more disordered, providing a possible explanation for the less regular
magnetic field structures observed in interacting galaxies such as M51 \cite{2010arXiv1001.5230F}.
%
%
\section{Conclusions}
We have shown that the contribution of CRs to the dynamics of the ISM on a global galactic scale, studied by means of CR+MHD simulations, leads to
a very efficient magnetic field amplification on the timescale of galactic rotation. The model applying a fixed analytical gravitational potential
reveals a large scale regular magnetic field with apparent spiral structure in the face-on view and a X-shaped structure in the edge-on view.
In the presence of spiral perturbations excited in the stellar component by a satelite galaxy, the magnetic field structure follows these perturbations in the stellar and gaseous components.
The magnetic field structure becomes less regular compared to the axisymmetric case. Dynamical magnetic field structures with opposite polarities develop within the disk and are present even at the saturation phase of the dynamo. Moreover, during the coalescence phase of the two galaxies the magnetic field structure becomes irregular as observed in M51.
An important part of the CR-driven dynamo is the galactic wind which reaches velocities of a few hundred km/s at galactic altitudes of a few kpc. The mass of gas transported out of the disk is about $1M_\odot/\yr$ for the star formation rate of the Milky Way.
\subsection*{Acknowledgements}
This work was supported by Polish Ministry of Science and Higher Education through the grants 92/N--ASTROSIM/2008/0 and N N203 511038.

\end{document}